\documentstyle[12pt]{article}
\begin{document}
\title{Black Holes From Diifferent Perspectives}
\author{B.G. Sidharth\\
Centre for Applicable Mathematics \& Computer Sciences\\
B.M. Birla Science Centre, Adarsh Nagar, Hyderabad - 500063 (India)}
\date{}
\maketitle
\begin{abstract}
In this paper we consider black holes from a non general relativistic
perspective as also from a microphysical point of view.
\end{abstract}
\section{Introduction}
It is generally believed that the concept of Black Holes requires
General Relativity for its understanding and interpretation. In this brief
note we will show that Black Holes could also be understood without invoking
General Relativity at all.\\
We start by defining a Black Hole as an object at the surface of which, the
escape velocity equals the maximum possible velocity in the universe viz.,
the velocity of light. We next use the well known equation of Keplerian
orbits\cite{r1},
\begin{equation}
\frac{1}{r} = \frac{GM}{L^2}(1+e cos \theta)\label{e1}
\end{equation}
where $L$, the so called impact parameter is given by, $Rc$, where $R$ is the
point of closest approach, in our case a point on the surface of the object
and $c$ is the velocity of approach, in our case the velocity of light.\\
Choosing $\theta = 0$ and $e \approx 1$, we can deduce from (\ref{e1})
\begin{equation}
R = \frac{2GM}{c^2}\label{e2}
\end{equation}
Equation (\ref{e2}) gives the Schwarzchild radius for a Black Hole and can be
deduced from the full General Relativistic theeory\cite{r2}.\\
We will now use (\ref{e2}) to exhibit Black Holes at three different scales,
the micro, the macro and the cosmic scales.
\section{Black Holes}
Our starting point is the observation that a Planck mass, $10^{-5}gms$ at the
Planck length, $10^{-33}cms$ satisfies (\ref{e2}) and, as such is a Schwarzchild
Black Hole (Cf.ref.\cite{r3}). Infact Rosen has used non-relativistic Quantum
Theory to show that such a particle is a mini universe\cite{r4}.\\
We next come to stellar scales. It is well known that for an electron gas in a
highly dense mass we have\cite{r5}
\begin{equation}
K\left(\frac{\bar M^{4/3}}{\bar R^4} - \frac{\bar M^{2/3}}{\bar R^2}\right)
= K' \frac{\bar M^2}{\bar R^4}\label{e3}
\end{equation}
where
\begin{equation}
\left(\frac{K}{K'}\right) = \left(\frac{27\pi}{64\alpha}\right) \left(\frac{\hbar c}
{\gamma m_P^2}\right) \approx 10^{40}\label{e4}
\end{equation}
and
$$\bar M = \frac{9\pi}{8}\frac{M}{m_P} \quad \bar R = \frac{R}{(\hbar /m_ec)},$$
$M$ is the mass, $R$ the radius of the body, $m_P$ and $m_e$ are the proton
and electron masses and $\hbar$ is the reduced Planck Constant. From
(\ref{e3}) and (\ref{e4}) it is easy to see that for $\bar M < 10^{60}$, there
are highly condensed planet sized stars.
(Infact these considerations lead to the Chandrasekhar limit in stellar
theory). We can also verify that for
$\bar M$ approaching  $10^{60}$ corresponding to a mass $\sim 10^{36}gms$, or
roughly a hundred to a thousand times the solar mass, the radius $R$ gets
smaller and smaller and would be $\sim 10^{8}cms$, so as to satisfy (\ref{e2})
and give a Black Hole in broad agreement with theory.\\
Finally for the universe as a whole, using only the theory of Newtonian gravitation,
it is well known that we can deduce
\begin{equation}
R \sim \frac{GM}{c^2}\label{e5}
\end{equation}
where this time $R \sim 10^{28}cms$ is the radius of the universe and $M \sim 10^{55}gms$
is the mass of the universe.\\
Equation (\ref{e5}) suggests that the universe itself is a Black Hole. It is
remarkable that if we consider the universe to be a Schwarzchild Black Hole
as suggested by (\ref{e5}), the time taken by a ray of light to traverse the
universe equals the age of the universe $\sim 10^{17}secs$ as shown elsewhere
\cite{r6}.
\section{The Kerr-Newman Formulation for the Electron}
It was already noted\cite{r7}, that a particle with the
Planck mass viz. $10^{-5}gms$ could be considered to be a Schwarzchild black
hole whose radius is of the order of the Planck length viz., $10^{-33}cms$. One
could then ask whether a charged rotating black hole, that is a Kerr-Newman
black hole could represent an elementary particle with charge. Indeed the
remarkable fact has been well known\cite{r2} that the purely classical Kerr-Newman metric does
describe the electron, including its purely Quantum Mechanical anomalous
gyro magnetic ratio $g=2$! This could have been construed to be the much sought
after unification of General Relativity and Quantum Mechanics, except for the fact
that such a Kerr-Newman electron black hole would have a naked singularity.
That is, its radius becomes complex:
\begin{equation}
r_+ = \frac{GM}{c^2} + \imath b, b \equiv (\frac{G^2Q^2}{c^8} + a^2 -
\frac{G^2M^2}{c^4})^{1/2}\label{e6}
\end{equation}
where $G$ is the gravitational constant $M$ the mass and $a \equiv L/Mc,L$ being
the angular momentum.\\
Even in the derivation of the above Kerr-Newman metric, Newman has noted\cite{r8}
the puzzling fact that an imaginary shift of coordinates has to be invoked
and that it is this imaginary shift which gives the rotation or spin. From
a classical point of view this is inexplicable.\\
On the other hand it has been pointed out by the author\cite{r9} that the
Quantum Mechanical coordinate of a Dirac electron is given by
\begin{equation}
x = (c^2p_1 H^{-1}t + a_1) + \frac{\imath}{2} c\hbar (\alpha_1 - cp_1 H^{-1})H^{-1},\label{e7}
\end{equation}
where $a_1$ is an arbitrary constant and $c\alpha_1$ is the velocity operator
with eigen values $\pm c$.\\
It has also been noted that for the electron the imaginary parts in (\ref{e6})
and (\ref{e7}) are of the same order, and that this imaginary coordinate was
given a physical explanation long ago by Dirac\cite{r10}: This has to do with the
famous Zitterbewegung. As space time intervals shrink, by Heisenberg's Uncertainty
Principle the uncertainty in the momentum - energy values increases. Thus only
averages over space time intervals, specifically at the Compton scale,
are physically meaningful. Within the Compton scale we encounter unphysical
Zitterbewegung effects which show up as complex (or non-Hermitian) coordinates
(Cf.also ref.\cite{r11}).\\
So the unsatisfactory feature of the Kerr-Newman electron black hole can be
circumvented with the Quantum Mechanical input that rather than space time
points as in classical theory, we need to consider averages over Compton
scale space time intervals.\\
All this pleasingly dovetails with the fact that minumum space time cut offs
can be taken consistently with the Lorentz transformation as shown by Snyder
a long time ago\cite{r12}. In recent years there has ben a
return to ideas of discrtee space time including through string theory\cite{r13}.
Further if the cut off is at the Compton
scale ($l,\tau$) then we have a non commutative geometry\cite{r13} viz.,
\begin{equation}
[x,y] = 0 (l^2), [x,p_x]=\imath \hbar (1-l^2)\label{e8}
\end{equation}
and similar equations, which can be shown to directly lead to the Dirac
equation. In other words it is this minimum Compton scale space time cut off
as in equation (\ref{e8}) which leads to the Dirac matrices, spin and
the anomalous gyro magnetic ratio $g=2$. Indeed it has been noted recently
by Ne'eman\cite{r14} that such a non commutative geometry provides a
rationale for renormalization. Infact this was the motivation for the very
early work of Snyder and others in introducing discrete space time.\\
It must be mentioned that if in (\ref{e8}) terms $\sim l^2$ are neglected,
then we recover the usual commutation relations of Quantum theory.\\
\section{Some Experimental Consequences}
The question that arises is, are there any experimental consequences of the
above formulation\cite{r15}:\\
I. We first observe that the magnetic component of the field of a static electron
as a Kerr-Newman black hole is given in the familiar spherical polar coordinates
by (Cf.refs.\cite{r9,r11}).
\begin{equation}
B_{\hat r} = \frac{2ea}{r^3} cos \Theta + 0(\frac{1}{r^4}), B_{\hat \Theta} =
\frac{ea sin\Theta}{r^3} + 0(\frac{1}{r^4}), B_{\hat \phi} = 0,\label{e9}
\end{equation}
whereas the electrical part is given by
\begin{equation}
E_{\hat r} = \frac{e}{r^2} + 0(\frac{1}{r^3}), E_{\hat \Theta} = 0(\frac{1}{r^4}),
E_{\hat \phi} = 0,\label{e10}
\end{equation}
A comparison of (\ref{e9}) and (\ref{e10}) shows that there is a magnetic
component of shorter range apart from the dipole which is given by the first
term on the right in equation (\ref{e9}). We would like to point out that a
short range force, the $B^{(3)}$ force, mediated by massive photons has
indeed been observed at Cornell and studied over the past few years\cite{r16}.\\
On the other hand as the Kerr-Newman charged black hole can be approximated
by a solenoid, we have as in the Aharonov-Bohm effect, a negligible magnetic
field outdside, but at the same time a real vector potential $\vec A$ which
would contribute to a shift in phase. Infact this shift in phase is given
by (Cf. also ref.\cite{r17})
\begin{equation}
\Delta \delta_{\hat B} = \frac{e}{\hbar} \oint \vec A . \vec{ds}\label{e11}
\end{equation}
There is also a similar effect due to the electric charge given by
\begin{equation}
\Delta \delta_{\hat E} = -\frac{e}{\hbar} \int A_0 dt\label{e12}
\end{equation}
where $A_0$ is the usual electro static potential given in (\ref{e10}). In the above Kerr-Newman
formulation, $(\vec A ,A_0)$ of (\ref{e11}) and (\ref{e12}) are given by (Cf.refs.\cite{r9,r11})
\begin{equation}
A_\sigma = \frac{1}{2}(\eta^{\mu v}h_{\mu v}),\sigma,\label{e13}
\end{equation}
From (\ref{e13}) it can be seen that
\begin{equation}
\vec A \sim \frac{1}{c} A_0\label{e14}
\end{equation}
Substitution of (\ref{e14}) in (\ref{e11}) then gives us the contribution of the
shift in phase due to the magnetic field.\\
II. Let us now consider some imprints of discrete space time, as discussed in
section 1.\\
First we consider the case of the neutral pion. As is known, this pion decays
into an electron and a positron. Could we think
of it as an electron-positron bound state also\cite{r11,r18,r19,r20}? In this case we
have,
\begin{equation}
\frac{mv^2}{r} = \frac{e^2}{r^2}\label{e15}
\end{equation}
Consistently with the above formulation, if we take $v = c$ from (\ref{e15}) we get the correct Compton wavelength
$l_\pi = r$ of the pion.\\
However this appears to go against the fact that there would be pair annihilation
with the release of two photons. Nevertheless if we consider discrete space time,
the situation would be different. In this case the Schrodinger equation
\begin{equation}
H \psi = E \psi\label{e16}
\end{equation}
where $H$ contains the above Coulumb interaction could be written, in terms
of the space and time separated wave function components as (Cf. also ref.\cite{r21}),
\begin{equation}
H\psi = E \phi T = \phi \imath \hbar [\frac{T(t-\tau)-T}{\tau}]\label{e17}
\end{equation}
where $\tau$ is the minimum time cut off which in the above work has been taken to be the Compton
time. If, as usual we let $T = exp (irt)$ we get
\begin{equation}
E = -\frac{2\hbar}{\tau} sin \frac{\tau r}{2}\label{e18}
\end{equation}
(\ref{e18}) shows that if,
\begin{equation}
| E | < \frac{2\hbar}{\tau}\label{e19}
\end{equation}
holds then there are stable bound states. Indeed inequality (\ref{e19}) holds
good when $\tau$ is the Compton time and $E$ is the total energy $mc^2$. Even if
inequality (\ref{e19}) is reversed, there are decaying states which are relatively
stable around the cut off energy $\frac{2\hbar}{\tau}$.\\
This is the explanation for treating the pion as a bound state of an electron
and a positron, as indeed is borne out by its decay mode.
The situation is similar to
the case of Bohr orbits-- there also the electrons would according to classical
ideas have collapsed into the nucleus and the atoms would have disappeared. In
this case it is the discrete nature of space time which enables the pion to be
a bound state as described by (\ref{e15}).\\
Another imprint of discrete space time can be found in the Kaon decay puzzle,
as pointed out by the author\cite{r22}. There also  we have equations like
(\ref{e16}) and (\ref{e17}) above, with the energy
term being given by $E(1 + i)$, due to the fact that space time is quantized.
Not only is the fact that the imaginary and real parts of the energy are of
the same order borne out but as pointed out in\cite{r22} this also
explains the recently observed \cite{r23} Kaon decay and violation of the time reversal
symmetry. In the words of Penrose\cite{r24}, "the tiny fact of an
almost completely hidden time-asymmetry seems genuinely to be present in the
$K^0$-decay. It is hard to believe that nature is not, so to speak, trying to
tell something through the results of this delicate and beautiful experiment."\\
From an intuitive point of view, the above should not be surprising because time
or even space reversal symmetry is based on a space time continuum and is no longer obvious
if space time were discrete.\\
Indeed this can be seen from equation (\ref{e8}): If we retain terms $\sim l^2$
then there is no invariance under not just time but also space reflections. It is within this framework
that we can also explain the handedness of the nearly massless neutrino: It has
a comparatively large Compton wavelength $l$ and by (\ref{e8}) space
reflection symmetry no longer holds.

\end{document}